\documentstyle[12pt]{article}
\title{Opto-electronic  application of AgInSe$_{2}$ }
\author{D. K. Shukla, Manohar Lal \thanks{Permanent Address:
G.G.D.S.D. College, Sector 32, Chandigarh. }  ~and Navdeep Goyal \\
{\it Centre of Advanced Study in Physics,\/}\\
{\it Panjab University, Chandigarh-160014 (India). \/} }
\date{}
\begin{document}
\maketitle
\begin{abstract}
The paper reports a possible application of AgInSe$_{2}$ for opto-electronic
switching. Material has been studied over a wide range of frequencies (5Hz to
1MHz), through measurements of conductance and capacitance, at different
temperatures and illumination levels. The results indicate that there is an
increase in capacitance (C) as well as conductance (G), when sample is exposed
to light radiations at a given temperature. The switching/recovery time has been
analyzed in terms of time constant ($\tau$ = C/G) and found to be of the order
of micro seconds for this material. It has been further observed that $\tau$
decreases with increasing illumination levels and temperature. It is
understandable, because higher the rate of recombination of optically/thermally
generated carriers, lesser should be the value of $\tau$.
\end{abstract}
\newpage
\section{Introduction}
\par The semiconducting I-III-VI$_{2}$ and II-VI compounds are of great
practical interest in fabrication of opto-electronic device. I-III-VI$_{2}$
compounds normally crystallize into diamond like structure and have attracted
considerable attention [1-3] of many workers. Most of these compounds can be
made to have both ${n-}$type and ${p-}$type conduction and are useful in the
field of photovoltaics [4,5]. Copper-III-VI$_{2}$ compounds have been studied by
many investigators [5-8]. The use of these compounds in the development of
photoconductive, photovoltaic and other devices has been established [6-8]. We
report in this paper some investigations made on a silver based chalcopyrite
semiconductor(AgInSe$_{2}$). AgInSe$_{2}$ is a material of special interest,
because it is a ternary analogue of CdS, with direct energy gap of 1.19 eV [1].
The study of AgInSe$_{2}$ communicated from the author's laboratory [9],
indicate the existence of gap states in this material. Present investigations
provide clear indication of suitability of AgInSe$_{2}$ for use as
opto-electronic switches. The switching time is of the order of microseconds for
planer AgInSe$_{2}$ samples.
\section{Experiment}
\par AgInSe$_{2}$ was prepared by sealing a desired quantity of
constituent elements (99.99\% pure) in an evacuated quartz ampule, which
was heat treated in a rocking furnace [9]. EDAX analysis of the prepared
material provides the following percentage compositions:
Ag: 24.982 At\%, In: 25.014 At\%, Se: 50.004 At\%.
The X-ray diffraction pattern of the sample shows that the material is
polycrystalline in nature \cite{r9}. The sample, in the form of pellets
was prepared by finely grinding the ingots and then compressing the
powder in a die under hydraulic pressure. The samples were annealed at
80 degree celsius for 24 hours.
Sheet measurements on planer configuration were made by exposing a
narrow slit of material ($\approx$ 1mm) in between two electrodes obtained by
coating a conducting layer of silver paste on one face of pellet.\par
Measurements were made on modular a.c. impedance and C.V system 
(EG \& G, PARC, USA) shown schematically in Fig. 1.
\par The real and imaginary components of a.c. impedance were
obtained over a frequency range 5Hz to 100kHz by using dual lock-in amplifier. 
Plots of capacitance versus time and  conductance versus time were
obtained at 1MHz on a C-V system  coupled with an X-Y recorder. The sample
was illuminated by using optical fibre bundle connected to Ealing (UK)
optical illumination set with a light source (Tungsten Halogen Sylvania
Lamp, 150W ) of variable intensity.
\section{Results and Discussion}
\par A.c. measurements are reported here, over a wide range of
frequencies (5Hz -1MHz) under dark and illuminated conditions for
AgInSe${_2}$. Figure 2. shows the variation of conductance with time at a
fixed temperature (T= 285 K) by switching the light ON and OFF for a
particular interval. It is evident from figure that there is an increase
in the conductance when sample is exposed to light radiations at a higher
temperature. Similar measurements were obtained on capacitance versus
time plot and it was observed that the capacitance of the sample
increases  when light radiations are shone on the sample. Similar
behaviour were observed at other temperatures, frequencies and
illumination levels. It is observed from the measurements under
illumination that there is a considerable increase in a.c.
conductance and the material is said to be in the ON (low resistance)
state. It goes back to OFF (high resistance) state, as soon as
illumination  is turned off. Figures 3 and 4 show the effect of
illumination on conductance and capacitance respectively. It is clear
from the figures that the material is sensitive to light over a wide range
of frequencies and the sensitivity of the material to illuminations
increases with the increasing frequency therefore, higher
frequencies are preferable for operations. The switching/recovery time
of the sample for OFF-ON-OFF states was estimated by determining the
RC-time constants ( ${\tau}$ = C/G) at different frequencies. Table 1
shows
\begin{table}[h]
\caption{Variation of time constant $\tau$ ($\times10^{-6}$ Seconds) at
different illumination levels (I$_0$=0.00, I$_1$=5.05, I$_2$=12.5,
I$_3$=26.5, measured in arbitrary units)}
\begin{center}
\begin{tabular}{|c|c|c|c|c|} \hline
\multicolumn{1}{|c|}{} &
\multicolumn{4}{c|}{Time constant $\tau$ at}\\ 
\multicolumn{1}{|c|}{Frequency} &
\multicolumn{4}{c|}{different illumination levels}\\ \cline{2-5}
\multicolumn{1}{|c|}{in kHz $\downarrow$} &
\multicolumn{1}{c|}{I$_0$} &
\multicolumn{1}{c|}{I$_1$} &
\multicolumn{1}{c|}{I$_2$} &
\multicolumn{1}{c|}{I$_3$} \\ \hline
1.00 & 0.129 & 0.102 &
8.44 & 6.12 \\ \hline 3.98 & 0.111 &
8.54 & 7.40 & 5.35 \\ \hline 10.00 & 9.40 &
7.25 & 6.80 & 4.50 \\ \hline 39.80 & 6.81 &
5.23 & 4.44 & 3.45 \\ \hline 100.00 & 5.04 &
4.02 & 3.36 & 2.62 \\ \hline
\end{tabular}
\end{center}
\end{table}
the value of $\tau$ at diffrent levels of
illuminations and frequencies. It is clear from the table that the
switching/recovery time (${\tau}$), is of the order of microseconds for
AgInSe$_2$. It is also clear from the table that switching response is
faster at higher frequencies. Figure 5 shows the frquency dependence of
time constant $\tau$ at different temperatures. It is clear from the
figure that the dependence of switching time on temperature decreases
at higher frequencies.
\section{Conclusion }
\par The response of AgInSe$_{2}$ to optical and thermal stresses
demonstrates the suitability of AgInSe$_{2}$ as an opto-electronic
switch over a wide range of frequencies. The switching/recovery time for
planer sample has been found to be of the order of microseconds.
\subsection*{Acknowledgments}
\par The authors are grateful to University Grants Commission for
providing funds under COSIST programme for purchase of a.c. impedance
and C.V. system. The authors are thankful to Professor K.K.Srivastava
for useful discussions.

\newpage
\listoffigures
\newpage
\begin{figure}
\vskip  12truecm
\caption{Schematic diagram of computer assisted a.c. impedance and CV
system set-up}
\end{figure}
\begin{figure}
\vskip  12truecm
\caption{Computer plot of conductance against time at T=285 K}
\end{figure}
\begin{figure}
\vskip  12truecm
\caption{Variation of conductance with illumination of sample
(illumination level 26.5 arbitrary units)}
\end{figure}
\begin{figure}
\vskip  12truecm
\caption{Variation of capacitance with illumination of sample
(illumination level 26.5 arbitrary units)}
\end{figure}
\begin{figure}
\vskip  12truecm
\caption{Effect of frequency on time constant $\tau$ at different
temperatures}
\end{figure}
\end{document}